\documentclass{article}

\usepackage{hyperref}
\usepackage{xcolor}
\usepackage{graphicx}
\usepackage{eso-pic}
\usepackage[misc,clock,geometry]{ifsym}

\usepackage{array, makecell} 
\usepackage{multirow}
\usepackage{epstopdf}
\usepackage{ulem}
\usepackage{caption}
\usepackage{footnote}
\usepackage{booktabs}
\usepackage{amsmath}
\usepackage{amsfonts}
\usepackage{amssymb}
\usepackage{upgreek}
\usepackage{thmtools, thm-restate}

\makesavenoteenv{tabular}
\usepackage{tikz}
\usepackage{epstopdf}
\usepackage{tablefootnote}

\usepackage{ulem}

\usepackage{algorithm}
\usepackage[noend]{algpseudocode}
\usepackage[affil-it]{authblk}

\usepackage{fontawesome}
\newcommand{\userYes}{\textcolor{black}{\faCircle}}
\newcommand{\userMaybe}{\textcolor{black}{\faAdjust}}
\newcommand{\userNo}{\textcolor{black}{\faCircleO}}
\newcommand{\LPYes}{\textcolor{red}{\faCircle}}
\newcommand{\LPMaybe}{\textcolor{red}{\faAdjust}}
\newcommand{\LPNo}{\textcolor{red}{\faCircleO}}

\title{SoK: Cross-Domain MEV}

\usepackage{authblk}
\author{Conor McMenamin\thanks{Email: \texttt{conor.mcmenamin1994@gmail.com} \\
This project was funded by the Uniswap Foundation Grants Program}}
\affil{Department of Information and Communication Technologies, Universitat Pompeu Fabra, Barcelona, Spain}

\usepackage{cancel}

\algdef{SE}[EVENT]{Hook}{EndHook}[1]{\textbf{Hook}\ #1\ \algorithmicdo}{\algorithmicend}\algtext*{EndHook}

\usepackage{pifont}

\def\bitcoin{%
  \leavevmode
  \vtop{\offinterlineskip %\bfseries
    \setbox0=\hbox{B}%
    \setbox2=\hbox to\wd0{\hfil\hskip-.03em
    \vrule height .3ex width .15ex\hskip .08em
    \vrule height .3ex width .15ex\hfil}
    \vbox{\copy2\box0}\box2}}

\newcommand{\newconstruct}[5]{%
  \newenvironment{ALC@\string#1}{\begin{ALC@g}}{\end{ALC@g}}
   \newcommand{#1}[2][default]{\ALC@it#2\ ##2\ #3%
     \ALC@com{##1}\begin{ALC@\string#1}}
   \ifthenelse{\boolean{ALC@noend}}{
     \newcommand{#4}{\end{ALC@\string#1}}
   }{
     \newcommand{#4}{\end{ALC@\string#1}\ALC@it#5}
   } 
}

\algdef{SE}[EVENT]{Upon}{EndUpon}[1]{\textbf{upon}\ #1\ \algorithmicdo}{\algorithmicend}\algtext*{EndUpon}

\begin{document}

\maketitle

\begin{abstract}
   We examine the current state of the cross-domain world, with particular focus on the protocols being used/planned for use by multi-domain users. We build on existing frameworks for analyzing how MEV is extracted, while also adding a new categorization of intrinsic-extractable value vs. time-extractable value to describe how MEV is generated for an extractor. Together, this provides us with a framework with which we compare classes of protocols enabling cross-domain MEV, and the MEV occurring within these classes. 
   
   We analyze each protocol class separately and compare current implementations to an ideal functionality for each. We primarily focus on analyzing the MEV mitigations that these protocols provide, both now, and into the future. In each case, we also outline the technical barriers that current protocol implementations face. With this methodology, we identify sequencers and order-flow auctions as cross-domain protocols with the greatest potential to mitigate MEV, but also as protocols with some of the biggest technical barriers. 
\end{abstract}

\section{Introduction}

The reality of a cross-domain world is fast approaching, and it may already be upon us. The latency wars that have occurred in other financial domains \cite{FlashboysOG,FlashBoys2.0} will come to cross-domain value extraction. To compete for MEV, there is an ever-increasing need for high-frequency traders to exist and extract on as many chains and domains as possible. While this is not necessarily a bad thing if there is a high degree of competition and decentralization, this game is typically winner takes all \cite{FlashboysOG}, which is definitely a bad thing. If one player controls access and pricing in the system, this player is free to charge arbitrarily high rents to use the system.
As such, there is a need to understand this cross-domain world, and proactively prepare for this oncoming reality. 

Previous work in the space has tried to define cross-domain MEV \cite{obadia2021unity} and hypothesize about how MEV changes in a cross-domain world. Despite the formalism in that work, it is unclear how to apply the provided formal definitions to draw qualitative or quantitative conclusions about cross-domain MEV. We note that \cite{obadia2021unity} is a self-proclaimed work-in-progress, and is likely intended to be thought-provoking rather than a standard for reasoning about cross-domain MEV in blockchain protocols. 

In this work, we systematize prominent cross-domain protocols, focusing on the types of MEV that occur within each protocol. We extend our analysis to protocols which can be used to capture and/or mitigate MEV in a cross-domain world, include protocols like batch auctions (Section \ref{sec:Batch}) and encrypted mempools (Section \ref{sec:encrypted}). Although not necessarily cross-domain enabling, we see these protocols as potential solutions to cross-domain MEV. As such, Section \ref{sec:where} analyzes protocols which either directly or indirectly affect  cross-domain MEV.

To do this, in each of the protocols, we breakdown the MEV occurring there by how the value is extracted, and where the value originates for the extractor. These categorizations are described in detail in Sections \ref{sec:how} and \ref{sec:whereFrom} respectively. By using a consistent set of MEV categorizations, we distinguish the types of MEV/methods of MEV extraction occurring in cross-domain enabling protocols, and what protections these protocols provide. As many of the most promising cross-domain protocols are still in the early stages of development, we also discuss the barriers that these protocols face.

Our analysis identifies the censorship-resistant ideal functionalities of shared sequencers (Section \ref{sec:sequencers}), multiple single-domain sequencers (Section \ref{sec:multisequencers}), off-chain requesting for quotes (Section \ref{sec:RFQs}), and batch settlement protocols (Section \ref{sec:Batch} as viable solutions to cross-domain MEV, and MEV in general. All of these protocols are not without technical barriers, as described in the respective sections, but stand as areas of particular interest for further research.

\section{Cross-Domain MEV $\subseteq$ MEV}\label{sec:Categories}

There are many potential categorizations for, and components of, MEV. We can consider extractable value based on how the value is extracted (e.g. $EV_{ordering}$ vs $EV_{signal}$ \cite{FrontierSignalvsOrdering}), where the value originates (Section \ref{sec:whereFrom}), who controls access the extractable value (e.g. Mafia vs. Monarch vs. Moloch \cite{3EV}), where the extraction occurs, how the extracted value can be isolated (e.g. proposer-builder separation (PBS, see Appendix \ref{sec:PBS}), order-flow auctions (see Section \ref{sec:OFAs}), etc.), who the extracted value can be redistributed to (users vs. builders vs protocols vs. everyone), to name but a few. Each of these distinctions has their own benefits and insights for each type of user within the system, be they protocol designers, searchers, users, block builders or validators. 

In this work, we focus on the initial two areas; how the MEV is extracted and where the extractable value is originating. Using these categories and components, we provide insight into the cross-domain MEV landscape, especially within the protocols that we introduce in Section \ref{sec:where}. 
We consider extractable value from the extractor's perspective. We note that extractor value may vary depending on ability to propose blocks, capital, latency, etc.. We generalize the notion of extractor to be the extractor with the most value to gain, as this is typically the extractor that wins the right to perform the extraction (at least in the PBS model). Furthermore, extractor value vs. user/protocol value is not necessarily zero-sum, because of the utility that users and protocols get from performing blockchain actions, making the sum of utilities positive.

\subsection{How Value is Extracted}\label{sec:how}

This categorization has already been developed through a series of blog-posts advocating it \cite{FrontierSignalvsOrdering}. To summarize, MEV is either extracted as a result of \textit{ordering} or \textit{signal}. Extraction due to ordering involves strategically inserting and sequencing a set of transactions to be executed, where the value of tokens at the start of the sequence is greater than the value at the end of the sequence, using on-chain data only. The clearest example of this is starting the sequence with some token $X$, and ending the sequence with a greater balance of token $X$, although it can be extended to tokens with the same peg (although this has its limitations \cite{USTDepeg,USDCDepeg}). Ordering extraction was popularized in the original Flashbots paper \cite{FlashBoys2.0} (signal extraction was not), and quantified here \cite{FlashbotsExploreWebsite}.

Signal extraction is based on information occurring off-chain which creates a profitable opportunity to submit a set of transactions on-chain. Signals can occur in the mempool, on other blockchains, centralized exchanges, in news headlines, to name but a few. A signal-based extractable value opportunity occurs when the value to act on some transaction or on-chain state for an extractor becomes higher as a result of new off-chain information. 

With a knowledge of underlying token valuations and/or the existence of hedging opportunities off-chain, signal opportunities involve the extractor starting and ending the on-chain extraction sequence in different tokens. It is with these valuations/off-chain opportunities that the starting token balance is worth less than the final balance to the extractor.

A common example of signal extraction is loss-versus-rebalancing \cite{LVRRoughgarden}, when the price of a token swap on a centralized exchange moves sufficiently far away from an on-chain AMM price, creating a profitable opportunity for an extractor to realign the AMM with the centralized exchange price. This signal opportunity is commonly known as CEX-DEX arbitrage. 

Both signal and ordering extraction techniques are used in blockchains. The profit from ordering for extractors, after paying for the right to extract, is almost 0 due to its risk-free nature (fork attacks aside) and its deterministic valuation (approximately the same profit for all extractors). Profit from signal extraction is much more subjective, depending on factors like risk-appetite, access and costs to trade against centralized exchanges, signal latency, etc.

\subsection{Where the Extractable Value Originates}\label{sec:whereFrom}

\begin{figure}
    \centering
    \includegraphics[scale=0.32]{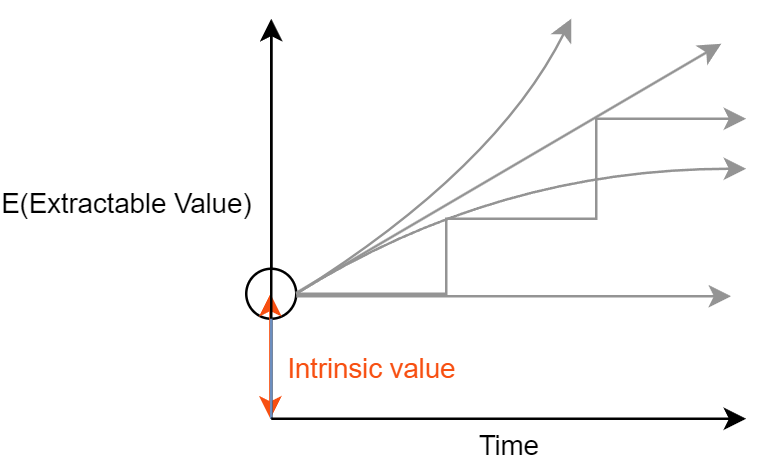}
    \caption{A representation of how the expected extractable value of transaction/protocol states with the same intrinsic-extractable value at expiration is increasing in the time until expiration. This is analogous to the time-value of options in traditional finance \cite{TimeValueOptions}. The exact expected extractable value function depends on the transaction/protocol, and the optionality that the extractor has. }
    \label{fig:TimeValue}
\end{figure}

This categorization has likely been hinted at in other works, although we have not seen it explicitly used. We propose that the expected extractable value of a transaction can be be split into mutually exclusive \textit{intrinsic-extractable value} and \textit{time-extractable value} components. This is exactly the same way options are valued \cite{OptionsPricing}. We note that due to extractor competition, the time at which positive extractable value opportunities can be acted on typically corresponds to the time at which the extractable value opportunity must be acted on. There are exceptions to this, particularly related to monopolization of inclusion/censorship powers. In what follows, we assume that the first time at which a state/transaction creates a positive extractable value opportunity corresponds to time when the state/transaction must be acted upon. We refer to this time as \textit{expiry/expiration}.\footnote{When censorship is possible, time-extractable value is bounded below by time-extractable value between now and the first time at which the state can be acted on.  However, expected-extractable value can decrease between state-update times (e.g. decreasing variations of the fee escalators introduced here \cite{FeeEscalator}). As such, generalized statements about the expected extractable value of a state increasing for an extractor over many state-update times are not possible.} 

\begin{itemize}
    \item \textbf{Intrinsic-extractable value} is the expected value to the extractor at the time when the blockchain state or transaction must be acted on.
    For a Uniswap order, this can be approximated by the maximum of the expected value of all of the front-running and back-running opportunities. For a Uniswap pool, it is the expected extractable value from moving the price up or down at the instant when orders are to be included in the blockchain (typically arbitraging the price in line with all other on- and off-chain liquidity).  Intrinsic-extractable value can be realized when the time to act on the order of blockchain state is 0 (the order/state must be acted on now). 
    \item \textbf{Time-extractable value} is slightly more complex. With respect to a blockchain protocol/transaction state on which the extractor can act on, the time-extractable value can be derived in a similar way to that of an option. The extractor has the length of time between confirmation times/blocks to decide whether or not to act on the blockchain state in question. The time-extractable value to the extractor of this optionality is the sum of all paths with a positive extractable value at expiration, times the probability of that path happening. \\
    For user transactions, time-extractable value is similar. For a transaction, the time value to an extractor is the expected value that can be extracted from the transaction between seeing the transaction and including the transaction. This is again analogous to options pricing. 
    \footnote{The intuition behind why time to expiration has a positive value for the extractor is the fact that the extractor has no obligation to include transactions at expiration. Furthermore, the extractor only includes transactions if the transactions' value to the extractor becomes positive at expiration. This creates a positive drift in the state's value to the extractor, with all negative paths removed from the sum, thanks to the optionality that the extractor receives. More information on why increasing the length of time to expiration increases the expected extractable value for an extractor, the owner of the option to act on the transaction/state, can be found here \cite{TimeValueOptions}.}
\end{itemize}

\subsubsection{Examples of MEV Opportunities Developing with Time}

A basic example of the value of time until expiration is the case of the prospector and the gold mine. A prospector is told by a surveyor that a mine has a $20\%$ chance to contain gold and be worth $\$1 $M, or an $80\%$ chance to contain no gold and be worth $\$0$. If the prospector is allowed to check the mine tomorrow and find out if it contains gold, how much would the miner pay for the chance to buy the mine tomorrow after inspecting it for $\$100$k? 
To calculate the value of the option, we can see that there is a $20\%$ chance that the miner makes $\$900$k (buys the mine), and an $80\%$ chance that the miner loses nothing, as there is no obligation to buy the mine. As such, this option is worth
\begin{equation}
    0.2* \$ 900\text{k} +0.8 * \$0= \$180\text{k}.
\end{equation}

In fact, as long as the cost to buy the mine is less than the value of the mine if it contains gold, the option to buy the mine has value. In this case, the current owner of the mine must charge for the option, or give away value to the prospector. With respect to blockchain protocols and transactions, this option is typically given away for fee. Although payoffs in the real-world are not so clear-cut, the same intuition applies. In the time between an extractor seeing an mempool transaction/protocol state and acting on it at expiration, the value of acting on the state changes\footnote{potentially millions of mini gold/no gold discoveries are made every second, in a phenomenon described by a Wiener process \cite{WienerProcess}}), but the cost for the extractor typically does not. 

A popular form of this cost which is actively and consistently being incurred by AMMs like Uniswap is loss-versus-rebalancing \cite{LVRRoughgarden}.\footnote{We note that censoring a protocol can also cause the value of the protocol to drop, potentially transferring the value to a competing protocol. Although this is an example of value being extracted due to time, it is not particularly useful for the analyses in this work.}
This effect is also felt by transactions in the mempool. Extractors and builders typically have no obligation to include transactions as they are submitted. Waiting the maximum amount of time at each blockchain update before deciding on whether or not to include a transaction allows extractors to maximize the expected extractable value from the transaction. This can be seen in the case where in the seconds before submitting a bundle/block, a significant event occurs in the real world which affects the underlying price of a token pair (like finding or not finding gold in the mine). By waiting until block-building time to decide on whether or not to include the transaction, the builder can choose to front-run the swap (profitable at block confirmation time) instead of back-running it (profitable when the transaction was first seen, but not profitable at block confirmation time).

\subsubsection{Using Intrinsic- vs. Time-Extractable Value to Analyze MEV}

When examining the MEV occurring in each of the protocols in the proceeding section, we consider the effects that these protocols have on both intrinsic- and time-extractable value. Intrinsic-extractable value can be reduced by things like reduced slippage requirements, batch execution, and/or encrypting some or all of a transaction's data. These are all things which restrict the price, execution capabilities, and/or information with which an extractor can act on transactions respectively. Users over-paying for a swap, execution freedom, and informational advantages are all things which allow an extractor to extract value from user transactions as soon as the transactions are seen; what we consider to be a transaction's intrinsic value.

Reducing time-extractable value is simpler in theory: reduce the time extractors have to act on a state, the time until expiration. As we will see in the proceeding section, this is fraught with complications. Another way to reduce the net time-extractable value paid by a user or protocol is charging for the time-extractable value that protocols and users give up (see Section \ref{sec:priority}), as is done in options trading.

From a recent result \cite{LVRandFeesRoughgarden}, the time cost for AMM pools grows proportionally to the square root of block time, which stands as the expiration time for the option to act on the AMM pool. This holds true in standard options pricing too, where the time value of options typically grows proportionally to the square root of time to expiration. For the purpose of this document, it suffices to understand that the time-extractable value for a protocol/transaction grows with respect to time until expiration, as in Figure \ref{fig:TimeValue}. As the time with which extractors have to act on a protocol or transaction approaches 0, so too does the time-extractable value, collapsing to the intrinsic-extractable value.

\section{Cross-Domain Protocols and MEV}\label{sec:where}

This section introduces a list of protocols at the core of cross-domain conversation. All of these protocols play some role in cross-domain MEV, with some protocols explicitly intended to provide some sort of MEV mitigation. In each of the protocols we discuss both the type of value extraction taking place (Section \ref{sec:how}), and where the extractable value typically originates (Section \ref{sec:whereFrom}). With each protocol we describe the ideal protocol functionality, current implementations, and open questions that exist towards implementing the ideal functionality specified. We also describe the state of MEV in each, including, but not limited to, cross-domain MEV. 

In the analysis that follows, we consider the centralization of power at any point in the transaction/block-building/MEV supply-chain, whether relaying, building, running an auction, block proposing, as negative for protocols. If protocols were to centralize power to a truly trusted player/set of players, this would fine. However, the existence of such players ad infinitum is unrealistic, and cannot be guaranteed in practice without strict rules and incentives. These rules and incentives do not exist as a given, and require careful construction.  In-keeping with the core decentralization values of blockchains,  we highlight centralizing effects that protocols have. Furthermore, in each such protocol, we leave the ``Can the protocol protect against such effects?'' as an important open question.

\begin{table}
\scalebox{0.62}{
\begin{tabular}{|l|c|c|c||c|c|||c|c||c|c|}
    \toprule
     & &  \multicolumn{4}{|c|}{\textbf{Current Implementations}} & \multicolumn{4}{|c|}{\textbf{Ideal Functionality}}  \\
    \midrule
     & & \multicolumn{2}{|c|}{\textbf{Origin}} & \multicolumn{2}{|c|}{\textbf{How Extracted}} & \multicolumn{2}{|c|}{\textbf{Origin}} & \multicolumn{2}{|c|}{\textbf{How Extracted}} \\
    \midrule
    \textbf{Protocol} & \textbf{Sec.} & \textbf{Intrinsic} &\textbf{Time} & \textbf{Ordering} &\textbf{Signal}&\textbf{Intrinsic} &\textbf{Time} &\textbf{Ordering} & \textbf{Signal} \\
    \midrule
    Multi-domain Sequencers & \ref{sec:sequencers} & \userMaybe \ \LPMaybe & \userMaybe \ \LPMaybe & \userMaybe \ \LPMaybe & \userMaybe \ \LPMaybe & \userNo \ \LPNo & \userNo \ \LPNo & \userNo \ \LPNo & \userNo \ \LPNo \\
    \midrule
    Single-domain Sequencers & \ref{sec:multisequencers} & \userMaybe \ \LPMaybe & \userMaybe \ \LPMaybe & \userMaybe \ \LPMaybe & \userMaybe \ \LPMaybe & \userNo \ \LPNo & \userNo \ \LPNo & \userNo \ \LPNo & \userNo \ \LPNo \\
    \midrule
    Order-flow auctions  & \ref{sec:OFAs} & \userMaybe \ \LPYes & \userMaybe \ \LPYes & \userMaybe \ \LPYes & \userMaybe \ \LPYes & \userNo \ \LPYes & \userNo \ \LPYes & \userNo \ \LPYes & \userNo \ \LPYes \\
    \midrule
    App-chains & \ref{sec:AppChains} & \userYes \ \LPYes & \userYes \ \LPYes & \userMaybe \ \LPMaybe & \userYes \ \LPYes  &  \userNo \ \LPNo & \userYes \ \LPYes &  \userNo \ \LPNo & \userYes \ \LPYes \\
    \midrule
    Priority Auction & \ref{sec:priority} & \userYes \ \LPMaybe & \userYes \ \LPMaybe & \userYes \ \LPMaybe & \userYes \ \LPMaybe & \userYes \ \LPNo & \userYes \ \LPNo & \userYes \ \LPNo & \userYes \ \LPNo \\
    \midrule
    Slot Auctions & \ref{sec:slots} & \userYes \ \LPYes  & \userYes \ \LPYes  & \userYes \ \LPYes & \userYes \ \LPYes  & \userYes \ \LPYes & \userYes \ \LPYes & \userYes \ \LPYes &  \userYes \ \LPYes  \\
    \midrule
    Batch Settlement & \ref{sec:Batch} & \userYes \ \LPYes & \userYes \ \LPYes & \userMaybe \ \LPMaybe & \userYes \ \LPYes &  \userNo \ \LPNo  &  \userNo \ \LPNo  &  \userNo \ \LPNo  &  \userNo \ \LPNo  \\
    \midrule
    Encrypted Mempool & \ref{sec:encrypted} & \userMaybe \ \LPYes & \userMaybe \ \LPYes & \userMaybe \ \LPYes & \userMaybe \ \LPYes &  \userNo \ \LPYes &  \userNo \ \LPYes &  \userNo \ \LPYes &  \userNo \ \LPYes \\
    \midrule
    Reduced Block Times & \ref{sec:reducedBlockTimes} & \userYes \ \LPYes & \userYes \ \LPYes & \userYes \ \LPYes & \userYes \ \LPYes & \userYes \ \LPYes &  \userNo \ \LPNo & \userYes \ \LPYes &  \userNo \ \LPNo \\
    \bottomrule
  \end{tabular}} 
     \caption{A breakdown of expected extractable value from users and LPs within an AMM protocol, when order submission to the protocol is controlled by the respective cross-domain protocols. The included protocols are not necessarily exclusive, and may be composed.\label{table:RelatedWorkComparison} The following labels* indicate that the corresponding MEV component: \\
  \userNo/\LPNo :\ Is effectively eliminated in the corresponding protocol for \textbf{users}/\textbf{\textcolor{red}{LPs}}. \\
  \userYes/\LPYes :\ Exists in the corresponding best-in-class protocol for \textbf{users}/\textbf{\textcolor{red}{LPs}}.\\
  \userMaybe/\LPMaybe :\ Is theoretically eliminated for \textbf{users}/\textbf{\textcolor{red}{LPs}}, but current implementations have seen limited deployment, depend on centralized trust, or significant technical barriers remain.\\
   *\footnotesize{Readers should consult the respective subsections for full details on the extent of these indicative labels, and the assumptions made related to the ideal functionalities.}}
  
\end{table}

\subsection{Framework for Analysis}

We consider blockchains as state machines\footnote{A state machine consists of set of variables, and sequence of commands/updates on those variables, producing some output.}. Blockchains can be represented as a directed acyclic graph, with each block containing state machine updates. The updates in a block $B$ act on the state achieved by applying all blocks in the directed sub-graph reachable from $B$. Where multiple competing sub-graphs exist, blockchain protocols have tie-breaking rules for deciding which sub-graph to choose (most blocks, greatest height, etc.). Unless otherwise stated, blocks contain a sequence of transactions (state machine updates) which must be applied in order\footnote{Transactions can also be batch executed as in Section \ref{sec:Batch}.}, with this sequence chosen by a single elected block proposer for each block\footnote{Protocols can force proposers to out-source sequencing to a dedicated sequencer (see Sections \ref{sec:sequencers} and \ref{sec:multisequencers}), or incentivize proposers to auction the right to sequence blocks (see Section \ref{sec:slots} and Appendix \ref{sec:PBS}).}. Blocks are added to the blockchain at discrete time intervals.

Where appropriate, our analysis focuses on the MEV occurring in a an AMM, where order submission is controlled by the respective protocols. We isolate the MEV being extracted from the users submitting orders for inclusion, and liquidity providers (LPs) deploying liquidity, in the case of AMM pools. This intended to simplify comparisons, as the action space is restricted to swap orders, and the value being extracted is intuitive. A summary of this analysis is provided in Table \ref{table:RelatedWorkComparison}.

\subsection{Shared Sequencer}\label{sec:sequencers}

\subsubsection{Ideal Functionality}

A single entity, the shared sequencer, accepts transactions for 2 or more blockchains, committing to include valid transactions in the order in which they are sequenced. Commitments can take place at any time (independently to block production time on the sequenced chains). The validity condition for transactions can range from fee payment, at a minimum, to `the transaction is executable in the given sequence'.

In an ideal implementation of a shared sequencer, the delay between valid transaction submission from users, and inclusion in the sequence of transactions is negligible. Furthermore, all builders are incentivized to obey the shared sequencer (for this to be true, builders would need to face economic punishment for disobeying the shared sequencer), and there is some way for builders to charge and control fees in the shared sequencer.
Moreover, an ideal shared sequencer allows for atomic cross-chain transactions, which are discussed in Appendix \ref{sec:atomictxs}. 

All of this practically removes extractable value opportunities. Intrinsic extractable value is prevented as no processing of transactions is possible, removing an extractor's ability to strategically bundle transactions. Time-extractable value is effectively eliminated as a result of the negligible confirmation times for transactions, which means no time for extractors to wait for more profitable opportunities. As soon as an extractable value opportunity exists with respect to a protocol, an extractor must take it, or miss the opportunity. 

Shared sequencing is intended to shift at least some of the signal MEV extraction to ordering MEV extraction (signals that previously existed across domains would exist on the same domain in the shared sequencer). If sequencing is first-come first-served, ordering MEV would be removed. As time-extractable value approaches 0, so too does the profitability of signal MEV. 

\subsubsection{Implementations and Discussion}

Prominent shared sequencer implementations include Espresso \cite{Espresso}, Astria \cite{Astria}, and Zenith \cite{Zenith}. Detailed write-ups on Espresso, Astria, and shared sequencers in general can be found here \cite{CharbonneauSequencers1,CharbonneauSequencers2}. All of these implementations are currently focused on sequencing layer-2s.

\subsubsection{Open Questions}

Unfortunately, shared sequencers are highly complex and depend on strong assumptions that are unlikely to hold without significant technological and incentivization advancements. Specifically, there is currently no protocol to guarantee that proposers and builders on sequenced chains obey the sequencer. This may be related to the difficulties in provably ensuring availability of data for all sequenced domains. Without proof that a sequence of transactions was delivered, punishments for not adhering to the sequence are limited. Data availability proofs on Celestia \cite{Celestia}, a leading data availability protocol, take upwards of 10 seconds, slower than almost all block production times. In the PBS paradigm that exists in Ethereum Proof-of-stake, relayers collect built block information, and are responsible for providing that information to the proposer and validators when a block is selected for inclusion. If proposers on sequenced chains must build blocks according to sequencer restrictions, the entire sequence needs to be available to the proposer before blocks are built. 

Related to centralization of power, the end-game for shared sequencers may be to also build the blocks for sequenced domains, much in the same way as proposed in SUAVE \cite{SUAVEIntro}(discussed in Section \ref{sec:OFAs}). This potentially shifts all of the MEV/transaction fee revenue control to the shared-sequencer, which does not align with the current paradigm of proposers on the individual chains controlling this revenue. 
A rational sequencer who can induce even small sequencing delays can extract all of the intrinsic extractable value from transactions. With centralization of the sequencer, such attack vectors are likely possible and undetectable. That said, time-extractable value is likely to be negligible in any sequencer implementation.

Existing shared sequencer solutions do not require transaction validity checks, which is problematic for a shared sequencer.  If a bundle of cross-domain transactions are intended to be all-or-nothing (all transactions are executed, or none), rational/malicious players only need to front-run one of the domains (on the shared sequencer) to invalidate the bundle. This in turn may invalidate all proceeding transactions in that domain. For a single domain where all-or-nothing restrictions are enforced more naturally, this is much less of an issue, if an issue at all.

There likely needs to be pre-processing or post-processing on shared sequencer transactions before the block is confirmed in which sequenced transactions are promised to be executable by one or more watchtowers. Incorrect (in)validation by a watchtower could then carry a severe penalty. These watchtowers can be executors or simply state-update checkers. There is a natural requirement therein to validate these transactions in the order they are sequenced. Although this is centralizing, a decentralized sequencer who can enforce punishments on these watchtowers neutralizes the bad effects it has (although if the validation is done before sequencing, we need competition. As such, post-processing is preferred, and can be done by even a single player, although we then need DOS protection).

\subsection{Multiple Single-Domain Sequencers}\label{sec:multisequencers}

\subsubsection{Ideal Functionality}

Each blockchain runs a sequencer, who commits to including valid sequenced transactions as they are sequenced. As in shared sequencers, this validity condition can range from fee payment, to executability. The delay between valid transaction submission from users, and inclusion in the sequence of transaction is negligible, with builders incentivized to obey the sequencer (for this to be true, builders would need to face economic punishment for disobeying the shared sequencer), and there is some way for builders to charge and control fees in the shared sequencer. Extractable value is eradicated in this ideal functionality, for the same reasons as the ideal functionality of the shared sequencer. Single-domain sequencers would not affect the signals existing outside of the chain being sequenced, although the signal MEV opportunities would reduce in line with the reduction in time-extractable value. 

\subsubsection{Implementations and Discussion}

A single-domain sequencer is exactly that; a sequencer in charge of a single blockchain. Arbitrum \cite{ArbitrumSequencer} and zkSync Era \cite{zksyncSequencer} are examples of blockchains using dedicated single-domain sequencers.
We have seen that shared-sequencer implementations are fraught with unsolved issues and considerations. In the short term, it seems we would be happier to accept the benefits of sequencing in a single domain at the expense of this added functionality. 

Furthermore, for a given transaction to be split across multiple domains, with single-domain sequencers, we can practically achieve much of the same functionality given the existence of cross-domain agents. Actions/intents requiring multiple domains can have conditional execution constraints controlled by the user, or delegated cross-domain executors.
As users can interact with directly with these cross-domain agents, the trust required is significantly less (pairwise economic/social/legal contracts can be established).

\subsubsection{Open Questions}

Current implementations of single-domain sequencers are centralized. Decentralizing the sequencer without degrading performance and usability is an active area of research. 
However, single-domain sequencers have significantly less technical barriers than shared sequencers, with connectivity and the alignment of incentives simplified in the single domain case. The trade-off here is the ability to execute cross-domain transactions atomically. As discussed above, this may not be such a big barrier.

\subsection{Order-flow Auctions}\label{sec:OFAs}

\subsubsection{Ideal Functionality}

Users are able to auction their orders/order information among a set of searchers. The proceeds of these auctions are returned to the user. An ideal cross-domain order-flow auction eliminates both intrinsic and time-extractable value for the user, assuming the order-flow auctions are indeed revenue maximizing. The revenue from such an auction should equal to total extractable value for the extractor.

Protocol-level value extraction (against the LPs) still remains. Multi-domain order flow auctions (like the proposed SUAVE protocol) will shift some of the signal MEV to ordering MEV in a similar fashion to shared sequencers.

\subsubsection{Implementations and Discussion}

The heavyweight contender in this domain is the Flashbots-proposed SUAVE (Single Unified Auction for Value Expression) chain \cite{SUAVEIntro}. Here, users can post transactions/intents to be executed across potentially many domains participating in the SUAVE domain. The general idea is for searchers (executors in the SUAVE notation) to compete via an auction to execute these intents, with some or all of the proceeds from this competition capturable by the user. The competition among searchers in SUAVE is intended to be based on partially encrypted transactions, with \textit{hints} provided by users revealing enough information to searchers to optimally bundle transactions, while retaining some form of pre-execution privacy, and the value that comes with it (extractable value that is not given to the searchers).

Searchers bundle transactions for inclusion in the SUAVE chain. The SUAVE chain logic then selects and orders the bundles maximizing the (extractable) value of the set of bundles according to the encoded valuation functions of the chain. These sets of bundles represent blocks to be added to participating blockchains. With enough participation, SUAVE is intended to act as a one-stop-shop for transaction submission across many blockchains. 

Uniswap have announced UniswapX \cite{UniswapXIntro}, an order-flow auction (Dutch Auction) to assist with routing in the proposed Uniswap V4 \cite{UniV4Release}. Due to the potential for thousands of liquidity pools for the same token pair in Uniswap V4, some off-chain routing is likely needed to ensure users receive the best possible pricing. Emulating a Dutch Auction among searchers to fill user orders should help in this regard. However, the UniswapX proposal also mentions the ability for auction winners to use their own private inventory/other on-chain liquidity, which might have negative implications for on-chain LP providers. A legacy order-flow auction protocol that appears to have shutdown in recent months is Rook \cite{RookWebsite}.

\subsubsection{Open Questions}

SUAVE depends on a centralized auctioneer \cite{SUAVEUpdate} (Flashbots or an additional trusted third party). The Flashbots hope is that this auctioneer can eventually be replicated by some form of trusted execution environment, and eventually a transparent decentralized system. The path to such an end-goal remains unsolved and a contentious point. Trust in a trusted auctioneer appears to be widely accepted, at least in the case of Flashbots as the trusted auctioneer, evidenced by the amount of blocks produced by MEV-boost \cite{MEVBoostpct}, which also depends on trusting Flashbots as an auctioneer.

The removal of extractable value through revenue-maximizing off-chain auctions depends on many factors, including searcher competition and auctioneer trust, both of which need further investigation. 
Without atomicity guarantees, most searchers in these domains must charge fees to off-set the cost and probability of not executing intended orders on dependent domains. Specifically, if a searcher agrees to sell some amount of tokens to a user in the order-flow auction with the intention of/assuming that they will be buying those tokens back on Ethereum, the searcher must increase their fee proportionally to the cost of not executing the swap on Ethereum. Sequencers would practically eliminate this cost. With such costs, searcher competition likely diminishes, reintroducing extractable value opportunities for dominant searchers.

The logic around maximizing extractable value through re-ordering bundles is intended to be run by trusted execution environments \cite{MillerSUAVETEE}, such as trusted off-chain hardware or cloud infrastructure.
Centralizing power off-chain to trusted execution environments requires significant further study, with guarantees/limitations of such solutions clearly stated for the community. 
Another concern for order-flow auctions is the ability to auction bundles of orders in a revenue-maximizing way for users generating the orders. The concern here is the creation of incentives to spam the auction and fill bundles with fake orders, paying less revenue to the affected users. This is discussed in detail here \cite{mazorra2023optimal}.

With respect to UniswapX, questions remain about the effect that improved user execution has for on-chain LPs. Given a particular Uniswap V4 pool is trading with a user, this pool must be offering a better price than any other on- or off-chain liquidity source, which also means the LP's price is likely mis-priced. To this end, some defragmentation of liquidity (restrict the execution of orders to Uniswap pools only) might be required to remove this potentially toxic element of efficient routing through off-chain auctions. 

\subsection{App-Chains} \label{sec:AppChains}

\subsubsection{Ideal Functionality}

A single instance of the protocol is implemented on a dedicated blockchain, with users required to bridge their assets to and from the chain. App-chains likely eliminate the intrinsic extractable value of orders, but their effect on time-extractable value is not clear. Like previous protocols merging users/order-flow from multiple domains, app-chains will also shift some of the MEV from signal to ordering, although signal opportunities likely still remain in the discrete block time, single proposer model. Based on the thesis of Osmosis \cite{Osmosis}, ordering extractable will be fully redistributable to both users and LPs. 

\subsubsection{Implementations and Discussion}

Related to shared sequencers is the idea of app-chains. An app-chain is exactly that, a separate blockchain/side-chain on which all app-specific actions are carried out. For a protocol such as Uniswap with fragmented liquidity, isolated/unused domains are exposed to increased time-extractable value, in addition to the reduced fees from not being actively used. An app-chain removes this inter-protocol staleness. An app-chain remains exposed to $t$-staleness where $t$ is the time between confirmations on the chain. For a protocol like Uniswap, the hope would be the app-chain would adapt to the core needs of the protocol, and favour short confirmation times to address the costs associated with time-extractable value. 
App-chains also have the potential to reduce intrinsic-extractable value existing in the 1 app per chain paradigm by defragmenting liquidity, and the reduced slippage/higher indicative price accuracy this would bring. 

Examples of app-chains include Osmosis \cite{Osmosis}, dydx \cite{dydx}, and ThorChain \cite{ThorChain}. A detailed discussion on the value proposition of app-chains is provided here \cite{VariantAppChains}. 

An interesting feature of Osmosis, one of the emerging app-chains in the DEX space, is their prospective ProtoRev feature \cite{OsmosisProtoRev}. ProtoRev effectively tries to realign all on-chain liquidity sources after each transaction, back-running every transaction with an attempted arbitrage of on-chain liquidity pools against each other. The surplus from this process (there can only be a surplus under current conditions) is shared among the DAO controlling the Osmosis chain. 
Arbitrary post-processing is not considered plausible in general purpose L1 chains, due to the added computational cost. That being said, ProtoRev demonstrates the power that app-chains have given an ability to tailor consensus and execution specifically to the needs of the dedicated application. 

\subsubsection{Open Questions}

The main problem that app-chains face is the need to bridge assets to and from the app-chain (which may have prohibitive costs, including additional order staleness, depending on the bridge), and worse user experience as a result. App-chains on their own do not necessarily affect the timing issues that exist on normal blockchains. Therefore, it is important to understand the timing, trust and security specifications of any proposed app-chain when considering its use. 
It has been demonstrated that app-chains can potentially provide meaningful ordering protections where multi-purpose blockchains cannot \cite{ChitraUncertainty}. This is in line with the promises of Osmosis. The extent to which these protections can be provided for each application stands as an open question.

\subsection{Priority Auction to Interact with Application First} \label{sec:priority}

\subsubsection{Ideal Functionality}

The ability to interact with the protocol is auctioned off, with proceeds paid to the protocol. This eliminates extractable value against LPs in expectancy, with the proceeds of both LP- and user-extractable value paid to the protocol itself. By explicitly capturing some of the MEV proceeds through the auction, net profitability is reduced, however, the gross MEV from ordering and signal techniques should remain largely unchanged.

\subsubsection{Implementations and Discussion}

A proposed implementation of this is MEV-capturing AMMs (McAMMs) \cite{McAMMs}, which proposes the protocol itself auctioning off the right to send the first swap to an Mc-AMM ahead of time, with reduced protocol performance if the transaction is not included. The auction proceeds should reflect the expected extractable value to be extracted from the pool within the block, both from the intrinsic extractable value of user orders, and the time-extractable value of both orders and the pool pricing function between blocks. In this regard, McAMMs have many similarities to slot auctions, although here the proceeds are more clearly exploited by the protocols themselves. 

Diamond \cite{DiamondMcMenamin}, a variation of a priority auction which shifts this auction to the builders, has both been described as a prospective Uniswap V4 pool \cite{DiamondV4}, following the release of Uniswap V4 specifications \cite{UniV4Release}. Builders must repay some percentage of the net tokens removed in the pool within the block. The final pool price in a block is claimed to correspond to the builders best guess of the true underlying price of the pool, with this net amount of tokens removed corresponding to the MEV as a result of time-value extraction. 
Comparing these two solutions, Diamond does not restrict access to any individual ahead of block production time, removing any possibility of denial-of-service attacks either on, or by, the player with the exclusive right to execute the first swap.  

\subsubsection{Open Questions}

Both of these solutions effectively require the builder/searchers to arbitrage the pool against this true price, and then provide liquidity to the user orders. In these solutions, the pool reserves play a limited role beyond providing users with an estimate of how their orders will be executed. On the other hand, the end game in PBS also appears to be for builders to implicitly provide liquidity to user orders, taking on inventory risk in order to extract all available value from a block. Importantly, McAMMs and Diamond return the value to where it is being extracted, as opposed to the block proposers in PBS, so has clear advantages over existing solutions in this regard.

Priority auctions likely incentivize private order-flow. By ensuring a significant share of order-flow, an extractor can better price priority auctions, allowing such an extractor to minimally increment the second best auction bid, guaranteeing priority. This creates a cycle, attracting more private flow. This in turn allows the extractor to provide private order-flow extracting arbitrarily high rents as the only player in town with access to the protocols offering priority. Such negative centralizing externalities need to be properly addressed.

Additionally, in the case where auctions must be carried out by a trusted auctioneer, such as in McAMMs, it is not clear if the incentives of the auctioneer are aligned with that of protocol. A clear specification and analysis of the role of the auctioneer is required.

\subsection{Slot Auctions}\label{sec:slots}

\subsubsection{Ideal Functionality}

Block proposers auction off the right to propose a block ahead of time. For a player winning multiple slot auctions across several chains, this gives the ability to simultaneously propose a block on these chains. This unlocks the ability to provide users with atomic cross-chain transactions (Appendix \ref{sec:atomictxs}). Atomic order-inclusion confirmation has the potential to eliminate time-extractable value if these confirmations are provided immediately, with sequence guaranteed. Retaining the ability to sequence transactions provides these slot auction winners with some time-extractable value (although less than if the transactions were not confirmed to be included). By simultaneously executing cross-chain transactions, signal MEV is reduced, while ordering MEV opportunities stand to increase, due to the increase in possible orderings for proposers/searchers controlling multiple simultaneous blocks. 

\subsubsection{Implementations and Discussion}

Related to shared sequencing is the idea of slot auctions, where block proposers auction off the right to build the block ahead of time. This idea was discussed in a recent blog-post \cite{alanaLevinSlotAuctions}, with the Interchain Scheduler \cite{InterchainSchedulerIntro} an initial proposed solution.

\subsubsection{Open Questions}

The exact implementation of such a system is yet to be released. As with any solution requiring builders to control and coordinate themselves across multiple domains, slot auctions centralize the builder set across domains. This is likely exacerbated with slot auctions due to the need to place bids ahead of time. Making the right to build in a particular slot tradeable via a slot market may mitigate this effect, allowing builders to adjust to information closer to proposal time. 
How this could be done is beyond the scope of this work. 

Incentivizing multi-chain searchers to win simultaneous slots across multiple chains creates a single point of failure shared by each of the chains. Requiring searchers to pay for blocks before receiving the transactions creates unstudied incentives related to private order-flow. Players with more private order-flow are better positioned to predict the value of a slot for each chain and able to pay the minimum amount to guarantee control of those slots. This centralizing vector requires further research.

\subsection{Batch Auctions/Batch Settlement}\label{sec:Batch}

\subsubsection{Ideal Functionality}

Orders are executed at a common price. In an ideal functionality with no censorship, extractable value opportunities are effectively eliminated. 

\subsubsection{Implementations and Discussion}

The most common example of such a batch settlement protocol is CowSwap \cite{CoWSwapWebsite}. In CowSwap, orders are exposed to time-extractable value due to the previously discussed power of inclusion that solvers in the protocol have when deciding which orders to execute in a batch. It can be argued that intrinsic value is reduced, especially when some trust exists in the solvers, as orders are less likely to be maximally extracted against in a batch.  Penumbra \cite{Penumbra} also proposes batch settlement, with orders there encrypted, mitigating extractable value. In Penumbra, trust is placed in the threshold-decryption committee to not decrypt information prior to inclusion in the batch. 

\subsubsection{Open Questions}

Transaction censorship in batch auctions diminishes the benefits of such protocols, exposing users orders to both intrinsic- and time-extractable value. Furthermore, searchers and/or block producers in charge of order inclusion can estimate the imbalance of an auction, either from unencrypted order information or from the implicit information leakage through identity-reveal, timing, etc.. These players can then use their 'last-look' capabilities to enter an auction if the implicit settlement price of a batch is profitable. This is time-extractable value, albeit diluted amongst orders in a batch. Such value can be seen as a combination of both ordering and signal MEV, although we consider it more of a signal opportunity. How to reduce trust/censorship capabilities of searcher/solvers in current batch protocols is an important open question.

\subsection{Encrypted Mempool}\label{sec:encrypted}

\subsubsection{Ideal Functionality}

Orders are committed to the blockchain before order information is revealed to the block producer. Orders are decrypted and executed at some point in the future. User-level extractable value is removed when no information about a user's transaction can be learned before it is executed. The effect on LP MEV is unclear. Although we would expect protocols to adjust to an encrypted mempool paradigm, in isolation (without batch execution for example) LPs remain exposed to value-extraction attacks, such as losses-versus-rebalancing. 

\subsubsection{Implementations and Discussion}

Secret Network \cite{SecretNetwork} and Penumbra \cite{Penumbra} are examples of protocols using encrypted mempools. For detailed discussions on the effects of encrypted mempools on MEV see \cite{EncryptedMempoolsQuintus,EncryptedMempoolsCharb}.

\subsubsection{Open Questions}

As mentioned previously, trust is currently placed in decryption committees not to decrypt transactions before committing transactions to the chain. Solutions without a committee \cite{DelayEncryptionBurdges} are promising, although significant work is still required to make these techniques feasible at scale.

When generic state updates are encrypted, their executability and computational requirements are not necessarily known. This leads solutions like Penumbra and Osmosis to restrict the action space of transactions. Implementing encrypted mempools outside of application-specific chains like the ones mentioned remains an area requiring further research.

\subsection{Reduced Block times}\label{sec:reducedBlockTimes}

\subsubsection{Ideal Functionality}

The consensus protocol is changed to allow for quicker block times, reducing both time- and signal-extractable value for LPs. In an ideal functionality without censorship, this also holds true for users. Intrinsic- and ordering-extractable value opportunities still remain for any delay between block times, which we assume to be the case. 

\subsubsection{Implementations and Discussion}
 
There is a nuanced, yet crucial difference between a sequencer that can provide fast confirmations, and increasing the throughput of a system through faster block times. The former is typically intended to provide cryptographic/economic guarantees between sequencer and the proposer/validators that when a block is added, the block will contain the transactions included in the sequencer. It is only the block contents that are applied to the state machine, so all of the security still rests on the proposer/validator. 

Faster block-confirmation times are being seen across many L2 chains, where the security lost to increased speed/centralization can be off-set be the security and oversight of a slower L1 blockchain.

\subsubsection{Open Questions}

Consensus layer changes must be carefully considered. Reduced block times come at the expense of security, with faster blocks increasing centralization, and incentivizing attacks such as forking the blockchain \cite{FasterBlockTimeIssue}. Faster confirmation at the layer-1 level has also lead to significant liveness issues \cite{LivenessAttack1,LivenessAttack2}.

\subsection{Bridges}

\subsubsection{Ideal Functionality}

Allows for the passing of messages between two blockchains. Messages on both the sending and receiving chain are subject to the constraints of the respective blockchains. Bridges are exposed to vulnerability attacks, which we deem to be intrinsic value to extractors, extracted through ordering attacks.

\subsubsection{Implementations and Discussion}

Bridges alone are not necessarily responsible for MEV related to decentralized exchange of tokens. They do however play an important role in cross-domain MEV. Bridges facilitate swaps that otherwise would not be possible, allowing users to transfer value on one chain to another. A basic example of this is locking tokens in a smart contract (bridge contract) on the source chain. This information is passed to the destination chain through a corresponding bridge contract, which creates/unlocks tokens tied to the tokens locked up on the source chain. Eventually, if users on the destination chain want to unlock the tokens in the source-chain bridge contract, the tokens must be locked/destroyed on the destination chain. When this happens, and is communicated back to the source chain, the source chain tokens can be unlocked. Implementations of bridges can be found in all Ethereum-native layer-2 protocols (Arbitrum, Optimism, zkSync, etc. \cite{L2Summary}), with cross-chain bridges summarized here \cite{BridgeSummary}.

\subsubsection{Open Questions}

An important result related to bridges is the fact that all messages (the proof-of-lock/-unlock in the toy example) between separate blockchains must be sent by a trusted third party/set of parties \cite{Zamyatin2019Communication}. Done naively, this is highly centralizing. These bridge contracts/third parties are the key-holders for the locked tokens on the source chain. By compromising the key-holders/vulnerabilities in how the key-holders interact with the bridge, extractors can drain bridge contracts. This creates a vector for MEV that does not exist elsewhere, which effects almost all of the proceeding protocols that use off-chain data. Examples of bridge attacks can be found here \cite{BridgeAttacks}.

\subsection{Off-Chain User-Liquidity Provider Interactions (RFQs)} \label{sec:RFQs}

\subsubsection{Ideal Functionality}

The user requests a price/market in a particular set of tokens from a set of liquidity providers. The liquidity providers respond with the required prices, to which the user can (or must) respond, typically within a certain time. In this ideal functionality, extractable value is removed. 

\subsubsection{Implementations and Discussion}

Most importantly, this protocol is not applicable for automated MMs, requiring active MMs who respond in real time to users. As a result, we relegate RFQs to the end of the section and exclude RFQs from Table \ref{table:RelatedWorkComparison}, which tries to isolate protocol effects on automated MMs. Despite this critical difference, requests for quotes (RFQs) allow liquidity providers to eliminate intrinsic extractable value, and account for time-extractable value by charging a dynamic fee. Implementations include ODOS \cite{ODOS}, Hashflow \cite{Hashflow}, Hop Exchange \cite{HOPexchange}, and WOOFi \cite{WOOFi}. Correctly implemented RFQs remove time-extractable value for the user by allowing users to instantly trade with quoted markets. Existing solutions side-step this with uncompetitive quotes/high fees, or by giving MMs the final say on whether to include a user transaction or not.

\subsubsection{Open Questions}

Can RFQ-style solutions be implemented on a blockchains to allow for near-instantaneous trades after quotes are provided? Sequencers, dedicated RFQ-chains and/or SUAVE may help in this regard.

\section{Conclusion}\label{sec:conclusion}

We provide a summary of cross-domain protocols, not only as a guide to understanding cross-domain MEV, but also as a resource for understanding the cross-domain world in general. Some of the most promising protocols for enabling a cross-domain world, such as shared sequencers and generalized order-flow auctions, are also the protocols with the most open questions. This highlights that the cross-domain world of today is still under construction, and will likely be very different to that of the future. This future will be shaped by the answers to the open questions that we have posed among others. As value is created in the cross-domain world, so to will value extraction opportunities. That being said, many of the ideal protocol functionalities of Section \ref{sec:where} effectively eliminate MEV, or allow for its redistribution.
Censorship and centralization are common barriers for many of the protocols that we discuss. These stand as areas that warrant particular focus from the research community. 
\section{Acknowledgements}

Thanks to Ankit Chiplunkar, Alana Levin, Alex Obadia, Bruno Mazorra, Federico Landini, Quintus Kilbourn, and Xin Wan for their insightful reviews and comments. 

\addcontentsline{toc}{section}{Bibliography}
\bibliographystyle{plain}
\bibliography{references,UFPrefs}

\appendix
\section{Proposer-Builder Separation}\label{sec:PBS}

Proposer-Builder separation (PBS) \cite{PBSIntro}, popularized by Flashbots' MEV-Boost \cite{MEVBoostpct}, involves the proposers of blocks, as elected by the blockchain consensus mechanism, auctioning off the right to build the block to a set of competing builders. PBS is simultaneously intended to remove the barrier to entry for block proposing, and censorship of transactions. 

In Ethereum proof-of-work, proposing and building were done by the same individual. Block-building has always been computationally intensive due to the need to store the entire blockchain's state, and simulate state over many combinations and permutations of prospective transactions. More recently, it has become extremely competitive due to the need to maximize extractable value from included transactions. By allowing proposers to out-source this work while still profiting from block-building due to the auction process, everyone can theoretically be a proposer. Furthermore, although block building is competitive and centralizing, block proposing remains relatively decentralized. By involving two unique individuals in the block proposing mechanism, the hope is that one of these individuals (typically the proposer) will prevent transaction censorship. From the proposers perspective, this can be done by adding constraints on blocks that the proposer will accept from builders. A comprehensive list of resources on PBS can be found here \cite{PBSAlternatives}.

\section{Atomic Cross-chain Transactions}\label{sec:atomictxs}

We briefly outline atomic cross-chain transactions, a prospective advancement with the potential to add significant utility (and MEV opportunities) to the cross-domain MEV space. An atomic cross-chain transaction allows for the simultaneous execution of transactions across multiple blockchains. More than this, atomic cross-chain transactions may even allow for the moving of assets among multiple blockchains. This would allow for instantaneously bridging assets and using them on destination chains, atomic cross-chain swaps, and potentially even flash-loans. Such functionalities would unlock a world of use-cases and value for users. 

That being said, any atomic cross-chain transaction solution requires each participating blockchain to trust in a third party external to the blockchain itself \cite{Zamyatin2019Communication}. This is a contentious issue, especially with regard to decentralization. To the best of our knowledge, no satisfactory solution to these strict trust requirements exists. However, the promise and potential of atomic cross-chain transactions, especially within protocols like shared sequencers, warrants some consideration. Protocols potentially allowing for atomic cross-chain transactions are discussed in Section \ref{sec:where}.

\end{document}